\documentclass[twocolumn]{IEEEtran} 

\newif\ifpdf
\ifx\pdfoutput\undefined           
   \pdffalse                       
\else
   \pdfoutput=1                    
   \pdftrue
\fi

\ifpdf
   \usepackage[pdftex]{graphicx}
\else
   \usepackage{graphicx}
\fi

\usepackage{epsfig}

\def\BibTeX{{\rm B\kern-.05em{\sc i\kern-.025em b}\kern-.08em
    T\kern-.1667em\lower.7ex\hbox{E}\kern-.125emX}}

\begin{document}

\title{A Multifunctional Processing Board for the Fast Track Trigger of
the H1 Experiment} 

\author{David Meer, David M\"uller, J\"org M\"uller, Andr\'e Sch\"oning,
Christoph Wissing 
\thanks{Manuscript received June 15, 2001. This work was supported
in part by the Swiss National Science Foundation under Grant No.
2000-056967.}
\thanks{D. Meer, ETH Z\"urich, Z\"urich, Switzerland}
\thanks{D. M\"uller, Supercomputing Systems, Z\"urich, Switzerland}
\thanks{J. M\"uller, Supercomputing Systems, Z\"urich, Switzerland}
\thanks{A. Sch\"oning, ETH Z\"urich, Z\"urich, Switzerland}
\thanks{Ch. Wissing, Universit\"at Dortmund, Dortmund, Germany}}

\markboth{IEEE Transactions On Nuclear Science, Vol. XX, No. Y, Month
2001}
{Meer, M\"uller, M\"uller, Sch\"oning and Wissing: A Multifunctional
Processing Board for the Fast Track Trigger of the H1 Experiment}

\maketitle

\begin{abstract}
The electron-proton collider HERA is being upgraded to provide higher
luminosity from the end of the year 2001. In order to enhance the
selectivity on exclusive processes a Fast Track Trigger (FTT) with
high momentum resolution is being built for the H1 Collaboration. The
FTT will perform a 3-dimensional reconstruction of curved tracks in a
magnetic field of 1.1\,Tesla down to 100\,MeV in transverse momentum.
It is able to reconstruct up to 48 tracks within 23\,$\mu$s in a high
track multiplicity environment. The FTT consists of two hardware
levels L1, L2 and a third software level. Analog signals of 450~wires
are digitized at the first level stage followed by a quick lookup of
valid track segment patterns.

For the main processing tasks at the second level such as linking,
fitting and deciding, a multifunctional processing board  has been
developed by the ETH Z\"urich in collaboration with Supercomputing
Systems (Z\"urich). It integrates a high-density FPGA (Altera APEX
20K600E) and four floating point DSPs (Texas Instruments TMS320C6701).
This presentation will mainly concentrate on second trigger level
hardware aspects and on the implementation of the algorithms used for
linking and fitting. Emphasis is especially put on the integrated CAM
(content addressable memory) functionality of the FPGA, which is
ideally suited for implementing fast search tasks like track segment
linking.
\end{abstract}

\begin{keywords}
HERA, H1 Collaboration, Trigger, Track Trigger, Processor Board,
Supercomputing Systems, DSP, FPGA, CAM, LVDS
\end{keywords}

\section{Introduction}
\PARstart{C}{ollisions} of 920\,GeV protons and 27.6\,GeV electrons
(positrons) are studied with the HERA accelerator at DESY. The H1
experiment is situated at one of the two interaction points, where
electrons and protons collide at a frequency of 10.4\,MHz. The H1
detector is described in detail elsewhere \cite{H1}. Electron-proton
interactions are triggered by a four-stage trigger (L1 -- L4), which
reduces the data rate to about 10\,Hz. The first level (L1) is a dead
time free hardware trigger with a decision time of 2.3\,$\mu$s. At
this step, trigger information is fully pipelined and the trigger rate
is reduced to about 1\,kHz. A refinement of the L1 decision is
performed by L2 within 23\,$\mu$s, which reduces the trigger rate to
about 200\,Hz. After a positive trigger decision on L2, readout is
started, which takes about 1\,ms. During this time, a negative trigger
decision on L3 can abort the readout of the detector. In the case of a
negative L3 trigger decision aborting the detector readout, a
considerable reduction of dead time can only be achieved if the L3
decision time is less than 100\,$\mu$s. After finishing or aborting
readout the trigger pipelining is restarted. Finally the data are
passed to a processor farm (L4), where events are fully reconstructed
within 100\,ms.

During an extended shutdown in 2000/2001 the HERA accelerator is
upgraded to gain sensitivity for rare processes by delivering a
fivefold increase in luminosity. Consequently higher interaction and
background rates are expected. Events with high momentum transfers
$Q^2 > 100\,\textrm{GeV}^2$ which are triggered by calorimeter based
signals can still be triggered with high efficiency after the upgrade.
For exclusive final states at low $Q^2$, where the background rate is
largest, an upgrade of the existing track trigger is necessary.

Therefore, the H1 collaboration decided to build a Fast Track Trigger
(FTT) \cite{FTT}, which provides trigger signals to the trigger levels
L1 -- L3. The FTT is able to reconstruct tracks of charged particles
with high resolution and to find particle resonances.

A multifunctional processing board has been developed by ETH Z\"urich
in collaboration with Supercomputing Systems (SCS) \cite{SCS} to solve
various different processing tasks at L1 and L2. After shortly
summarizing the FTT system a detailed description is given about this
multifunctional processing board.

\section{The Fast Track Trigger (FTT)}
\begin{figure}[hbt]
  \begin{center}
    \ifpdf
       \includegraphics{fig1.pdf}\\
    \else
       \includegraphics{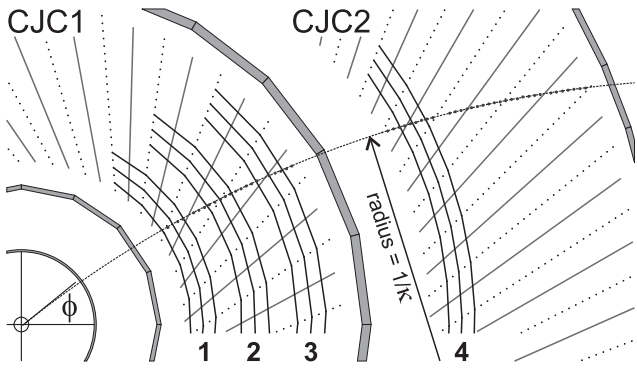}\\
    \fi
  \end{center}
  \caption{$x$-$y$ view of a charged particle track in the central jet
  chambers (CJCs). A track segment is defined by a group of three
  layers of hit wires. There are in total four of such trigger groups.
  A track segment is described by the azimuth~$\phi$ and the
  curvature~$\kappa$.}
  \label{cjc}
\end{figure}

The input of the FTT is based on charge and time information of the
inner central jet chamber (CJC1) with 24~radial wire layers and the
outer central jet chamber (CJC2) with 32~radial wire layers. Trigger
signals are built from 3~groups of 3~selected wires each in CJC1 and
one group in CJC2 as shown in Fig.~\ref{cjc}.
A track segment is defined by a three layer coincidence matching a
predefined hit pattern of vertex constrained tracks. Track segments
are described by the curvature $\kappa=\frac{1}{R}$, the
azimuth~$\phi$ and the declination~$\theta$.

The main task of L1 is to find track segments and to make a trigger
decision based on coarsely linked tracks. At L2 the track segments are
linked to tracks. The accuracy of track parameters is improved by a
3-dimensional track fit. After a positive trigger decision, these track
parameters are passed to the L3 processor farm, where the event is
fully reconstructed also taking into account other detector
information.

The FTT can process up to 48~tracks per event, which is fully
sufficient for 98\% of all events of interest.

\subsection{Finding Track Segments at L1}
\begin{figure}[hbt]
  \begin{center}
    \ifpdf
      \includegraphics{fig2.pdf}\\
    \else
      \includegraphics{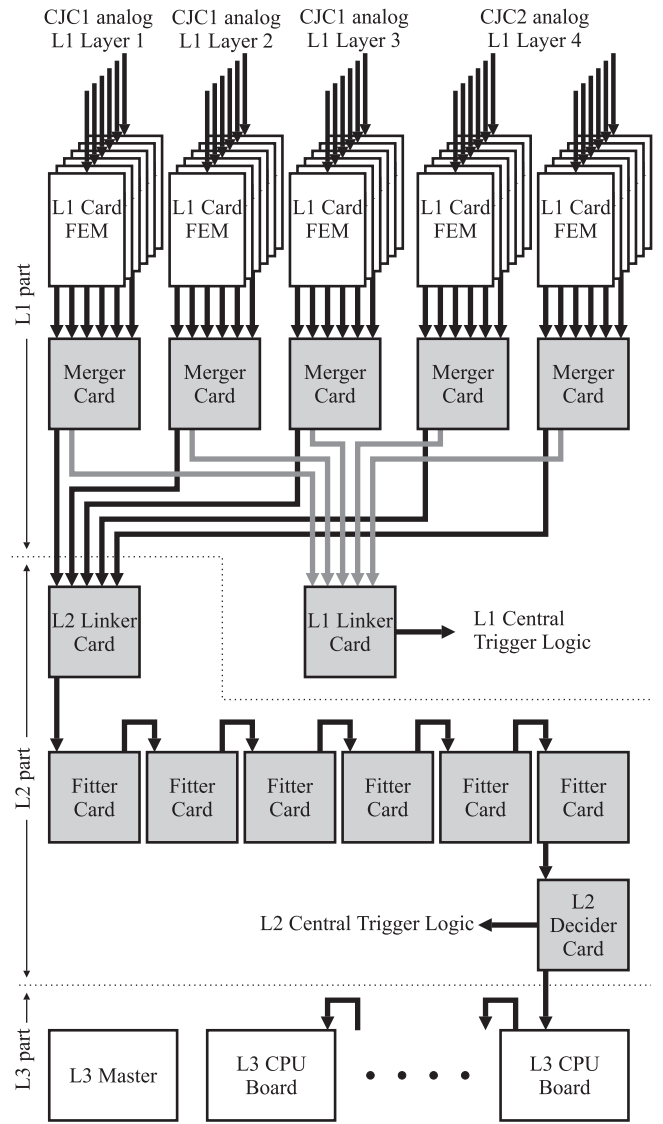}\\
    \fi
  \end{center}
  \caption{Hardware implementation of the FTT. The multifunctional
  processing board ({\sl shaded}) is used as merger card, L1/L2 linker
  card, fitter card and L2 decider card.}
  \label{fttoverview}
\end{figure}

Analog CJC signals of the existing readout system are tapped by
adapter cards and sent to the Front End Modules (FEM), see
Fig.~\ref{fttoverview}.
Signals are digitized at 80\,MHz using an 8-bit linear FADC and are
fed into shift registers. A farm of Field Programmable Gate Arrays 
(FPGAs) from Altera (APEX 20K400E) searches for predefined track
segment hit patterns. Track segment parameters are looked up from a
list of valid masks stored in SRAMs. Finally, they are sent via merger
cards to the linker cards for track linking. For a fast trigger
decision coarse linking is performed on the L1 linker card while the
L2 linker card exploits full resolution.

\subsection{Track Linking and Fitting at L2}
On the L2 linker card, track segments from the four trigger groups are
linked to tracks exploiting full resolution. The track linking is
performed by a fast and highly parallel algorithm which searches for
matching track segment parameters.

Track segments assigned to single particle tracks are sent to a total
of 6 fitter cards, where Digital Signal Processors (DSPs) perform a
3-dimensional helix fit. After fitting, track parameters are sent to the
L2 decider card, where a trigger decision is formed based on
kinematical or topological track quantities.

\subsection{Searching for Particle Resonances at L3}
A processor farm at L3 will reconstruct the event and search for
particle resonances also in high multiplicity events. The farm
consists of up to 16~commercial CPU VME boards (MVME 2400) equipped
with a 450\,MHz PowerPC750. Studies have shown that one processor
board has sufficient processing power to search for specific final
state topologies or decay channels within 100\,$\mu$s.

\section{The FTT multifunctional processing board}
For the integration of the various different processing tasks of the
FTT into hardware a multifunctional processing board has been
developed by ETH Z\"urich in collaboration with SCS. This board may
function as a L1 and a L2 linker card, as a fitter card and as a L2
decision card. The same board is also used as a merger card to connect
the large number of Front End Modules of the L1 system to the two
single L1 and L2 linker cards. The multiple use of the same board
design for different tasks considerably reduces development and
production costs. Depending on the main purpose of the board,
expensive components like high density circuits (FPGA, DSP) may be
omitted if not required.

\subsection{Design Overview}
\begin{figure}[hbt]
  \begin{center}
    \ifpdf
      \includegraphics{fig3.pdf}\\
    \else
      \includegraphics{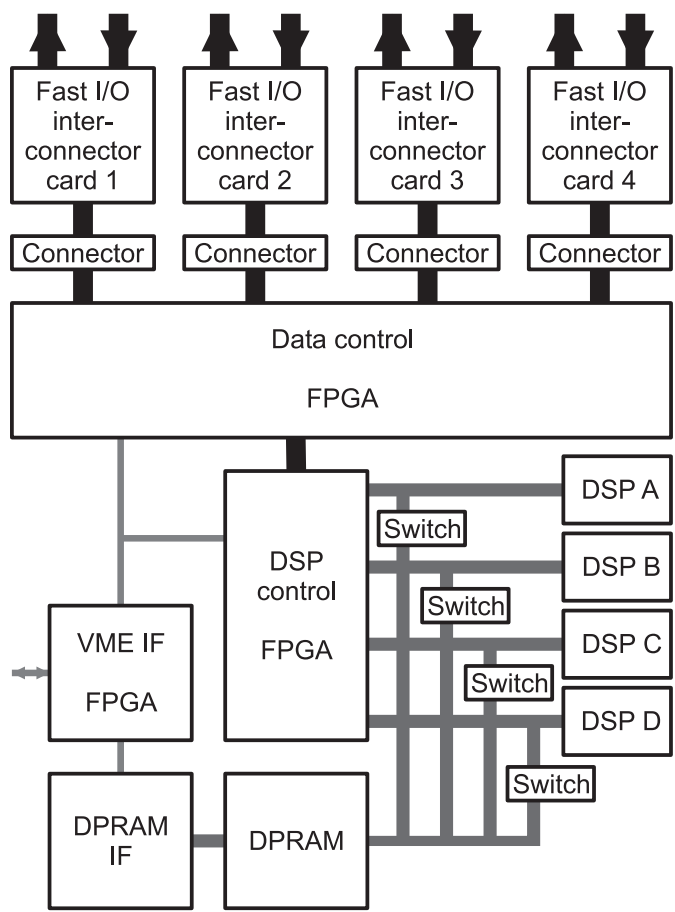}\\
    \fi
  \end{center}
  \caption{Functional block diagram of the multifunctional board.}
  \label{l2block}
\end{figure}

The multifunctional processor board is a compound of a main board
equipped with four DSPs and two large FPGAs and up to four I/O
interconnector cards (so-called ``Piggyback'' cards) which serve as
fast I/O interface between multifunctional processor boards. Two of
them are plugged on the top and two are plugged on the bottom of the
main board. A block diagram of the multifunctional board is shown in
Fig.~\ref{l2block}.

\subsubsection{Mainboard}
The main board is a 14-layer PCB board fitting in a double height (6U)
VME crate. For VME access, data distribution and monitoring via the
backplane, the board is equipped with a DIN96 VME connector and a
user-defined, metric connector.

The core of the L2 main board is formed by the four floating-point
DSPs (Texas Instruments TMS320C6701) \cite{TI} and a large FPGA
(Altera APEX 20K600E) \cite{Altera}. Mathematical algorithms, like
track fitting, run on DSPs while the large FPGA is predestined for
complex and associative logic like the linking of track segments. In
addition, the FPGA serves as an I/O data controller with bidirectional
connections to all fast I/O interconnector cards and is also connected
to a second FPGA (Altera APEX 20K200E) which serves as controller for
the four DSPs. To extend the internal 64\,KB DSP RAM for memory
extensive applications (e.g. lookup table) an external SRAM of 512\,KB
is connected to each of the DSPs. A third, smaller FPGA (Altera FLEX
EPF10K30A) is used as a VME interface. Data controller, DSP
controller, VME interface and a dual ported RAM (DPRAM) are connected
to a local bus. At startup, the DSP code is loaded via the VME
interface into the DPRAM, from which each DSP can download its own
program code by using a switch. Since multi-processor busses are
poorly supported by the TMS320C6701, each of them is connected to the
controller via an individual bus.

\subsubsection{Fast I/O Interconnector (Piggyback) Cards}
For data transmission between different multifunctional processing
boards a high-speed LVDS channel link is used. This 48-bit wide link
runs at about 104\,MHz thus providing a total data throughput of
$\rm 5.0\, Gb/s$. The I/O cards are equipped with a LVDS transmitter
(National DS90C387) and a LVDS receiver (National
DS90CF388)~\cite{National}. A small FPGA (Altera APEX 20K60E) serves as
controller and data switch between the LVDS input, the LVDS output and
the bidirectional connection to the main board. The main tasks of the
FPGA are the buffering of data coming from the different inputs and the
data distribution. The priorities for the receiving and the sending of
data can be programmed in a flexible way depending on the application.

A few Piggyback cards have the transmitter replaced by a second
receiver. This is required for the merger cards since in total six
LVDS input streams per merger board are needed.

\subsection{Clocking}
Both FPGAs, the data controller and the DSP controller, run at a
frequency of 104\,MHz. This frequency can either be generated from a
local oscillator or be derived by clock multiplication from the
10.4\,MHz HERA clock signal which has the frequency of electron-proton
collisions. The 104\,MHz clock is also conducted to the I/O
interconnector card and may be used for data transmission by the LVDS
channel link. In both cases it is essential to have a high-quality
clock with small jitter because a 7-fold multiplexing of up to
728\,MHz is used internally by the LVDS channel link and even a small
jitter would impair the data transfer quality.

There are independent clock domains from the receiving side of the
LVDS channel link running asynchronously to the local mainboard clock
at 104\,MHz. Incoming signals are buffered and synchronised in an
asynchronous FIFO on the I/O interconnector card.

Another clock domain of 41.5\,MHz is required for the DSPs. This clock
is generated by a local oscillator. The required frequency of 166\,MHz
for the DSP is derived by an internal phase-locked loop in the DSP.

The last clock domain is formed by the local bus running at 10.4\,MHz.

\subsection{Power Supply}
All devices of the main board use the LVTTL standard of 3.3\,V with
exception of some VME devices which are operated at 5\,V TTL. The core
of the APEX FPGAs needs 1.8\,V. These voltages are supplied via the
customized backplane. The 1.9\,V for the DSP core is generated
directly on the main board from the 3.3\,V.

\subsection{Communication}
The data transfer between different multifunctional processor boards
is realized by a messaging system which is based on routing tables.
Dynamic routing tables increase the flexibility of the FTT system and
ease maintenance. In the case of an extension or modification of the
system, the programming of the remaining cards can be kept and only
the routing tables have to be modified. In that scheme every
programmable component (FPGA, DSP) is able to send, receive or forward
a message to other components. In the current FTT implementation
(Fig.~\ref{fttoverview}) about 500 different messages are needed, each
having a so-called channel number assigned. The channel number is
represented by the first 9 bits of a 48-bit word. The remaining bits
may be used to transmit data, i.e. track segment information, as well
as control words. A routing table in every programmable component is
used as a lookup to send messages to an intermediate or final
destination. Individual routing tables are generated for all
programmable components so that transfer delays of messages are
minimized. The routing tables are implemented in internal RAMs and
contain a dynamic and a static partition, the latter being loaded when
configuring the board. The static partition already allows a minimal
communication between boards and the VME interface. Afterwards the
dynamic partition of the routing table is written during startup.

\section{Software and user code}
In the following the application-specific software of the
multifunctional processing boards is discussed with emphasis on the
track finding and track fitting algorithms to be implemented in the
programmable devices.

\subsection{Merger Cards}
The main purpose of the merger cards is the collection of track
segments from the six different FEM inputs, the multiplexing of data,
and the forwarding to one of the two linker cards (L1/L2). The user
code is rather simple and is not further discussed here. Since the
main task is the buffering of data in FIFOs, the equipment of the
mainboard with DSPs is not required.

\subsection{L1 Linker Card}
The main purpose of this card is the fast track linking and triggering
for L1. Input data to the L1 linker card are track segments from the
four radial trigger groups in the CJC as identified by the 30 FEMs and
are forwarded by the merger cards. The track segments are filled into
four corresponding, coarsely binned $\kappa-\phi$ histograms. The
histograms are stored in registers of the large data controller FPGA.
Track segment matches are defined by a coincidence of at least two out
of four trigger groups. The search is performed in all bins of the
histogram simultaneously.

A peak finder algorithm takes into account track segments from
adjacent bins as well. Finally the track multiplicities above momentum
thresholds and the track topology (e.g. two back-to-back tracks) are
used to form a trigger decision on L1. The complete linking algorithm
written in VHDL requires about 10000 FPGA logic cells using a
histogram of size $8 \times 60$ bins. A DSP is not used here.

\subsection{L2 Linker Card}
\begin{figure}[hbt]
  \begin{center}
    \ifpdf
      \includegraphics{fig4.pdf}\\
    \else
      \includegraphics{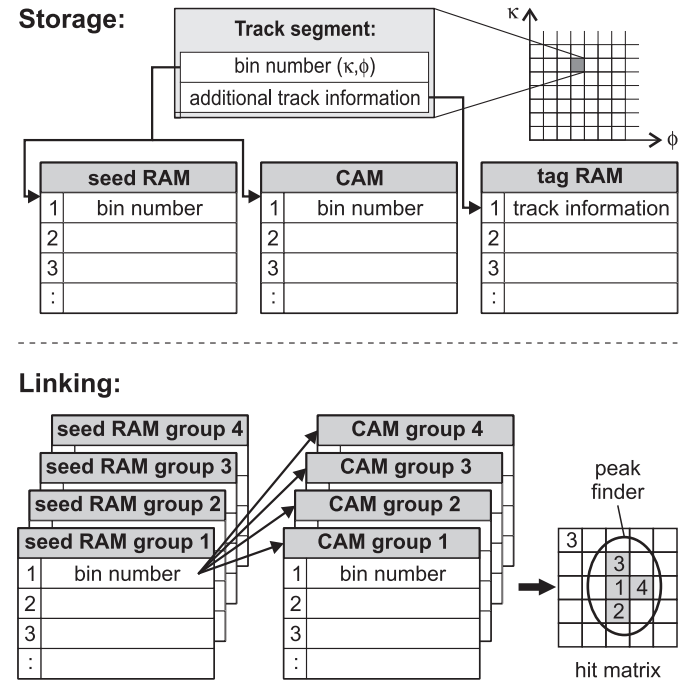}\\
    \fi
  \end{center}
  \caption{Storage and linking of track segments at L2. See text for
  details.}
  \label{l2linker}
\end{figure}

In contrast to the track linking at L1, where time requirements are
most stringent, the main constraint of the track linking at L2 comes
from the high-resolution requirement. The track segment storage and
linking is schematically shown in figure~\ref{l2linker}.
Similar to L1, the algorithm for track linking is implemented in the
data controller FPGA. However, in contrast to the L1 linker card,
received track segments are written into arrays (containing the
$\kappa-\phi$ locations in a ``virtual'' histogram) rather than stored
directly into a real histogram. The virtual histogram is divided into
40~bins of $\kappa$ and 640~bins of $\phi$ and has a much higher
resolution compared to L1. It is obvious that the standard way of
filling a histogram having about 50000 bins is not appropriate for
track segment storage and search. To solve that problem so-called
Content Addressable Memories (CAMs) are used.

A CAM can be regarded as inverse RAM where the input patterns are
compared with pre-loaded values and matches are indicated by the
corresponding address location in a single step. In particular, the
combination of a CAM with a tagged RAM (one-to-one correspondence of
addresses) allows a simple and condensed implementation of a large
lookup table. That combination makes it possible in a very efficient
way to store the non-zero entries of a large histogram with 50000 bins
and to search for specific entries in a single step without the need
for running a loop and sequential processing. This combined CAM and
RAM functionality, which is ideal for the implementation of any search
task in general, can be embedded very efficiently in the APEX 20KE
family of FPGAs.

In our example the CAM is filled with the addresses of the track
segment location in the histogram, the $\kappa-\phi$ bin. Additional
information, i.e. track segment parameters, are stored in the tagged
RAM. In addition, the track segment locations are filled into four
so-called lists of seeds which correspond to the four different trigger
groups. By running a loop the list of seeds is worked through. The
track segment locations are read and presented to the CAMs. If track
segments with identical histogram locations are found in at least two
trigger groups, a track segment link is defined.

In order to take into account migration effects between bins due to
the limited detector resolution, bins adjacent to the seed location
are also considered. The precise peak position of a cluster of linked
track segments is found by running a peak finder algorithm which is
based on a $3 \times 3$ sliding window and maximizes the number of
matched track segments. To be able to perform the two tasks (track
segment search and peak finding) in a highly parallel manner several
CAMs are installed per trigger group. In total 100 CAMs are
implemented such that all track segments in a $5 \times 5$ array
around a track segment seed in each of the four radial layers can be
searched through in parallel. The peak finder algorithm is highly
parallel and basically needs only one step to define a good track
link.

The complete algorithm (i.e. the receiving of data, the filling of the
CAMs and RAMs, the cluster finding around seeds and the peak finding)
is fully pipelined and runs at 104\,MHz. The linker card is designed
to link up to 48 tracks, which are afterwards distributed to a system
of six daisy-chained fitter cards.

\subsection{Fitter Cards}
The fitter cards perform helix fits of tracks to gain precision. A
single DSP on a fitter card starts as soon as all linked track
segments of the same track have arrived. In a first step, a
non-iterative circle fit \cite{Karimaki} in the $r$-$\phi$ plane is
performed by constraining the track to originate in $x$ and $y$ from
the primary vertex position of the electron-proton collision. The
circle fit takes about 330 clock cycles. In a second step, a fit in
the $r$-$z$ plane\footnote{$r$ is the radius and $z$ is the coordinate
along the beam axis} is performed, which determines the declination of
the track and takes about 200 clock cycles. The primary vertex
position in~$z$ is provided by the Central Proportional Chamber of the
H1 detector via a special interface card. After fitting, the final
track parameters are sent via the daisy-chained LVDS channel link to
the L2 decision card. For the FTT system it is foreseen that one DSP
performs up to two track fits per event.

\subsection{L2 Decider Card}
All tracks are collected and track-based quantities are evaluated to
form the L2 trigger decision. These quantities can be track
multiplicities above thresholds, momentum sums or simple topological
criteria (jets). The reconstructed tracks are processed in the data
controller FPGA within the remaining about $2.5\rm \, \mu s$ of the L2
latency. Exclusive final states, like particle resonances, may be
identified using the DSPs. Simulations have shown that it is possible
to calculate invariant masses of all two track combinations in an
event with low track multiplicity $N_{\rm tracks} \le 5$.

A positive L2 trigger decision is sent via the user defined backplane
to the H1 central trigger. In the case of a positive trigger decision
all track parameters are sent to the Power-PC farm of the FTT~L3
system for further event processing.

\section{Timing and performance}
The L2 latency of the H1 Trigger of 19.7\,$\mu$s gives a strict upper
limit on the FTT available time for generating a trigger decision. An
overview of the estimated timing at L2 is given in
Table~\ref{tab:timing}. All values are considered to be conservative for
an event with the maximum number of 48 tracks. First tracks are expected
to have finished fitting already after 8.5\,$\mu$s. Therefore, an
interleaving of tasks (e.g. start fitting before all tracks are
linked) will considerably reduce the overall processing time.
\begin{table}[htb]
  \caption{Estimation of the overall timing at L2 under the assumption
  that one DSP performs two track fits. An interleaving of tasks (e.g.
  fitting during linking) is not taken into account and would reduce the
  L2 latency.}
  \begin{center}
    \begin{tabular}{|l||c|c|}\hline
      Task     & Time $[\mu$s$]$ & Cumulated $[\mu$s$]$ \\ \hline \hline
      Latency L1-L2         & 0.404 &  0.404 \\ \hline
      Linking: receive data & 2.462 &  2.865 \\ \hline
      Linking: fill CAM/RAM & 0.096 &  2.962 \\ \hline
      Linking: check CAMs   & 5.115 &  8.077 \\ \hline
      Latency daisy-chain   & 1.413 &  9.490 \\ \hline
      Data Delay Fitting 1  & 0.501 &  9.991 \\ \hline
      Fitting 1             & 3.193 & 13.184 \\ \hline
      Data Delay Fitting 2  & 0.501 & 13.685 \\ \hline
      Fitting 2             & 3.193 & 16.878 \\ \hline
      L2 Decider card: Sums & 2.500 & 19.378 \\ \hline
      Spare time            & 0.322 & 19.700 \\ \hline
      Central Trigger       & 1.000 & 20.700 \\ \hline
    \end{tabular}
  \end{center}
  \label{tab:timing}
\end{table}

\section*{Summary}
A multifunctional processing board has been presented for the first
and second level system of the new Fast Track Trigger, which is built
for the H1 experiment. The system fulfills the hardware and timing
requirements and is able to reconstruct up to 48 tracks with high
resolution. It integrates several tasks like merging of data, track
linking, track fitting, and triggering, implemented in a single board
design with FPGAs and DSPs. The multifunctional processing board can
be flexibly utilized by using up to four high speed I/O interconnector
cards.

\nocite{*}
\bibliographystyle{IEEE}


\end{document}